\documentclass[12pt, letterpaper]{article}

\usepackage{amsmath,amssymb}
\usepackage{cite}
\usepackage{fancyhdr}
\usepackage[top=1in, bottom=1in, left=1in, right=1in]{geometry}
\usepackage{graphicx}
\usepackage{hyperref}

\numberwithin{equation}{section}
\date{}

\begin{document}
\title{{\rm\footnotesize \qquad \qquad \qquad \qquad \qquad \ \qquad \qquad \qquad \ \ \ \ \ \                      RUNHETC-2025-25
}\vskip.5in    Note on Type $III_1$ Algebras in $ c= 1$ String Theory and Bulk Causal Diamonds}
\author{Tom Banks\\
NHETC and Department of Physics \\
Rutgers University, Piscataway, NJ 08854-8019\\
E-mail: \href{mailto:tibanks@ucsc.edu}{tibanks@ucsc.edu}
\\
\\
}

\maketitle
\thispagestyle{fancy} 

\begin{abstract}  We argue that the Leutheusser-Liu procedure of isolating a von Neumann algebra in the $N = \infty$ limit of string theories, leads to the algebra of relativistic fermion fields on a half line for the $c = 1$ string theory. This is a Type $I$ von Neumann algebra, since it is the algebra of the Rindler wedge in the Rindler vacuum state. Subalgebras of finite regions are Type $III_1$. The argument uses the elegant results of Moore and of Alexandrov, Kazakov and Kostov.  This model is well known to be integrable and have no black hole excitations.  We have speculated that adding an interaction invisible in perturbation theory to a large finite number, $M$, of copies of the model, produces a non-integrable model with meta-stable excitations having all of the properties of linear dilaton black holes.  The algebra of fields is the tensor product of $M$ copies of the $c = 1$ model's algebra, whether or not we add the non-integrable interaction.  We argue that the infinite dimensional $c = 1$ algebras are analogous to those of the boundary field theory in AdS/CFT, even though they appear to encode bulk causal structure. An IR cutoff on the boundary renders them finite and causal structure must be formulated in terms of an analog of the Tensor Network Renormalization Group.  This is a time dependent Hamiltonian flow, embedding smaller Hilbert spaces into larger ones.  It is the analog of one sided modular inclusion in quantum field theory.  
\normalsize \noindent  \end{abstract}


\newpage
\tableofcontents
\vspace{1cm}

\vfill\eject
\section{Introduction}

In a recent series of papers\cite{LL}, Leutheusser and Liu have proposed an algebraic approach to studying local regions of space-time in the $N = \infty$ limit of AdS/CFT models.  They argue that in the actual $N = \infty$ limit the algebraic nature of the algebra of ``single trace" boundary operators, acting on the thermofield double state, can change from Type I to Type $III_1$ as a function of the temperature.  This is the sign of the appearance of a black hole horizon in the bulk geometry at the Hawking-Page transition.  

Another peculiar property of the limiting algebra is that the algebra of a finite boundary time interval in the vacuum sector, has a non-trivial commutant in the full operator algebra on Hilbert space.  LL associate that commutant with the algebra of operators in a causal diamond that just touches the causal wedge of the boundary in that interval.  

The procedure of isolating an algebra of single trace operators is applicable to any large N matrix quantum mechanics.  In this note we apply it to the double scaled limit of the Type 0B matrix model\cite{0B} which has a clear space-time interpretation, a connection to string theory, and is exactly soluble\cite{moore}.  AdS/CFT models have two dimensionless parameters, the ratio of the AdS radius to the Planck length, which is some power of a large integer $N$, and the ratio of the Planck scale to the string scale, which defines the string coupling $g_s$.  The $0B$ model only has one parameter, the ratio of the Fermi energy to the ``frequency" $\omega$ of the upside down oscillator.  When the Fermi energy, as measured by this dimensionless parameter, is far below the top of the potential, string perturbation theory is a valid approximation.  Roughly speaking, this model corresponds to AdS/CFT models with AdS radius of order the string scale.  Some models of the latter type are also exactly soluble CFTs, with no black hole excitations.
The strong coupling limit of the 0B model occurs when the Fermi energy approaches the top of the oscillator potential.  In AdS/CFT models with an Einstein-Hilbert dual, this corresponds to a regime where the classical gravity approximation is valid, but this is not true for the Type 0B string.  The low energy effective field theory of the Type 0B string is a version of linear dilaton gravity\cite{CGHS} with only two "Ramond - Ramond" fields.  That model has black hole solutions, but the exact quantum mechanical solution of the double scaled matrix model has no excitations with the properties of black holes.  Instead it is completely integrable and has an infinite number of conservation laws incompatible with the folklore derived from Hawking's analysis of black hole decay.  

The analysis of\cite{CGHS} shows that it's not surprising that classical physics fails for the Type 0B EFT, because validity of the classical field equations near the black hole horizon requires a large number of Ramond-Ramond fields.  The basic problem is that although the model allows for arbitrarily high entropy states, they all exist, as we'll see, in the asymptotic region where the coupling goes to zero. In\cite{lindil1} we proposed to remedy this problem by coupling together a large number $M$ of Type 0B models, with a coupling that was concentrated near the top of the potential.  The coupling was of four Fermi type and consisted of two pieces.  The first was a singlet under the $U(M)$ symmetry of the models and attractive.  It guaranteed that if one sent in a state such that a large fraction of the fermion flavors reached the top of the barrier at the same time, they would trap each other in a meta-stable bound state.   The second part of the interaction was an SYK like random coupling, which broke all symmetries of the problem.  The new model has the following properties:
\begin{itemize}
\item To all orders in perturbation theory, because the interaction is localized at the top of the potential, the model looks like $M$ decoupled copies of the integrable $0B$ string theory.
\item It has high entropy meta-stable excitations whose decay violates all the conservation laws of the perturbative expansion.  Since the violations of complete integrability are localized near the top of the potential we can get a qualitative understanding of the physics even in the strong coupling regime where it is relatively easy to excite the meta-stable states.  In this regime all of their properties are characterized by a single energy scale, the curvature at the top of the potential, if the SYK couplings take generic values from the random distribution.  In particular, the spectrum of emitted particles is thermal, with a temperature of order this scale.  Furthermore, if a new particle is thrown into an existing meta-stable state before it decays, the time for it to equilibrate is independent of the entropy and depends only on the same microscopic energy scale.  These are the same properties one can derive from the geometry of linear dilaton black holes.  All of these qualitative properties are true for large finite $M$.  The meta-stable boundstates do not have sharp horizons because they decay, with lifetimes of order their finite entropy.  

\end{itemize}

Thus, although the emergent infinite dimensional operator algebras are indeed related to the causal structure of a relativistic space-time, in this model they are not related to the properties of black holes, or horizons.  Indeed, we will argue that they are more like the boundary algebras of the AdS/CFT correspondence, whose bulk interpretation has to do with infinite volume divergences of the global spatial slices of AdS space.  The analog of a UV cutoff on the CFT, replacing it by a finite lattice approximation, is to replace the double scaled limiting theory by some finite $N$ matrix model with a deep double well potential, and a fermi surface close to but not infinitesimally close to the maximum between the two wells.  The fact that the depth of the Fermi sea is correlated with large values of the oscillator coordinate, which is mapped by AKK into the asymptotic region of Minkowski space, is a UV/IR duality analogous to Maldacena's scale/radius duality in the AdS/CFT correspondence. The mapping of ground state fermion entropy density in eigenvalue coordinate to dilaton field in the bulk gravity theory gives us a way to define a sequence of models, analogous to the Tensor Network Renormalization Group\cite{tnrg}, which define finite causal diamonds in this cutoff framework.

\section{Operator Algebras in the Double Scaled 0B Matrix Model}

The single trace operators in single matrix QM are traces of powers of the matrix.  The usual map into fermionic eigenvalue quantum mechanics converts these into fermion bilinears in second quantized notation.  Moore's careful treatment of the double scaling limit tells us that the limiting operator algebra is just the algebra of smeared fermion fields of the upside down oscillator problem, with smearing functions that have appropriate falloff at infinity.   In order to relate this to something easy to characterize, we use the transform of Alexandrov, Kazakov and Kostov\cite{AKK}, explained lucidly in\cite{ms}, which maps this into the algebra of a pair of massless Dirac fermions on a half line, with a scattering matrix
\begin{equation}   \Psi_1^{(out)} (r, t) = \frac{1}{\sqrt{2\pi}} \int_{-\infty}^{\infty} du e^{\frac{1}{2} (r + u)} [{\rm exp}(i\sigma_3 e^{r + u}) +{\rm exp}( \sigma_2 e^{r + u})] \Psi^{(in)} (u,t) .   \end{equation}  Here, using the notation of \cite{ms} $\Psi^{(out)}$ is a doublet of two component Majorana Weyl fermions, while $\Psi^{(in)}$ is a doublet of the opposite chirality.  The Pauli matrices act on the internal indices.  
In the Hilbert space based on the usual Minkowski vacuum, the Majorana field algebra is Type $III_1$ on a half line.  However the Hilbert space of the AKK transformed double scaled matrix model is more like that of the Rindler vacuum in ordinary QFT.  The scattering states are pure states\footnote{I thank S. Leutheusser and H. Liu for pointing this out to me.}.  The algebra is Type I in this space, as the notation of\cite{ms} makes clear.  Any restriction of the algebra to a finite region of the linear dilaton space-time is Type $III_1$.  

The algebras of the models made from $M$ copies of the $0B$ model are just finite tensor products of the $0B$ algebras and so share the same Murray-von Neumann type. 
More importantly, the algebra type does not change when we add the interactions at the top of the potential, which destroy integrability and produce meta-stable excitations.  So, unlike what happened in the $N = \infty$ limit of AdS/CFT models, the MvN classification of neither the full algebra of observables, nor that of finite regions is sensitive to the existence of black hole like excitations.  

\subsection{JT Gravity Coupled to Fermions}

In\cite{bdz} we used the results of\cite{moore} and\cite{AKK} to model massless fermions propagating in the throat of an extremal Reissner-Nordstrom black hole of large charge.  This is described by Jackiw-Teitelboim gravity, but unlike most of the literature,  we chose the background solution with negative coefficient for the dilaton field.  Since the fermions couple only to the AdS metric, which is conformally flat, we can use the AKK mapping to transform them into non-relativistic fermions in an upside down oscillator potential.  In this model only half the potential is used, or equivalently we restrict attention to wave functions of fixed parity.  The top of the potential is identified with the asymptotic boundary of $AdS_2$, from which the fermions reflect, while the region at infinity maps to the horizon of the patch where the dilaton field is positive.  SYK like four fermi interactions near the infinity of the eigenvalue coordinate\footnote{This model has only a single fermion field, so the interactions refer to couplings between different localized modes of this field.} guarantee that the system thermalizes and a state of the non-relativistic fermions is identified\footnote{This identification is done using semi-classical methods to evaluate traces over the single fermion Hilbert space.  There is no small parameter that justifies this approximation.}  which matches the fermion entropy in a region to the value of the classical dilaton field in JT gravity in causal diamonds.  Again, the mapping to a relativistic QFT allows us to isolate Type $III_1$ subalgebras of operators associated with causal diamonds.  

As in the linear dilaton models it's clear that the origin of infinite dimensional Hilbert spaces in these models has to do with the potentially infinite entropy associated with infinity in the upside down oscillator potential.  If we think of these models as originating from finite charge RN black holes, we know this must be cut off, and in\cite{bdz} we used the finite matrix model as the appropriate finite entropy cut off.  As we will show below for the linear dilaton model, this does not prevent us from defining sharp causal diamonds, although it does introduce ambiguity into their definition.  They are not completely determined by the truly asymptotic variables.  

\section{Tensor Network Models, Operator Algebras and Black Holes}

It is common in the literature on AdS/CFT to refer to tensor networks as ``toy models" of the correspondence.  This is inaccurate.  Instead they should be viewed (as they were originally by condensed matter theorists) as a sequence of lattice approximations to the CFT whose accuracy increases as one approaches the boundary of hyperbolic space.  This viewpoint is most explicit in the Tensor Network Renormalization Group\cite{tnrg} (TNRG) of Evenbly and Vidal, which has been shown to provide a systematic computation of critical exponents for simple $1 + 1$ dimensional models.  Fischler and the present author\cite{tbwfads1} argued that the embedding maps of the TNRG should be viewed as maps of the degrees of freedom in one causal diamond into those of the next larger one in a nesting of diamonds along a timelike geodesic in AdS space.  They define a discrete causal time evolution along a particular time-like geodesic in AdS space. The time intervals in the nesting are determined by the relative entropies of the diamonds, as measured by the areas of the corresponding shell of the network.  In this correspondence, individual nodes of the network are viewed as spheres of size $\sim R_{AdS}$, making explicit that the network is not local on sub-AdS radius scales.  The empirical evidence from the huge collection of extant AdS/CFT models suggests that the physics of these nodes also includes 2 or more compact dimensions, which have linear sizes of order the AdS radius.  

This construction makes explicit the scale-radius duality of Maldacena, the relation between proper time and radius in AdS geometry and, as Harlow\cite{harlowecc} has shown, incorporates the Quantum Extremal Surface and Ryu-Takayanagi formulae, as well as the qualitative properties of black holes.  Aspirationally, one would want to mimic\cite{happy} and always construct tensor networks invariant under discrete subgroups of the AdS group that map one network node into another.  Although it is unlikely that we will ever actually construct such a network for a model with an Einstein-Hilbert dual, it seems extremely plausible that they exist and provide some kind of definition of finite area causal diamonds for finite $N$.  Each shell of the tensor network is the holographic screen of a causal diamond along a particular timelike geodesic. Regularizations of quantum field theory are usually highly non-unique and only become universal for quantities that converge to finite limits when the cutoff goes to infinity, so this definition of finite diamonds is unlikely to have any canonical form, even if we impose as much symmetry as possible on it.  

This non-uniqueness is consistent with the observation\cite{tbsober} that local observers cannot access most of the quantum information in finite area causal diamond.  In the tensor network construction this is obvious for diamonds whose area is large in AdS radius units.  A single node of the network simply does not have enough q-bits to measure and store the data about the shell that surrounds it.  So local physics has neither a sharp mathematical definition, nor can it be probed by experiment, with infinite precision.

Let us be quite explicit about why this procedure succeeds in defining sharp causal diamonds without introducing infinite dimensional algebras.  The reason is that the time evolution defined by the TNRG is not generated by a time independent Hamiltonian.  Indeed, in the prescription described in\cite{tbwfads1} it only defines a sequence of unitary embeddings of smaller Hilbert spaces into larger ones.  The hypothesis of a TNRG invariant under a discrete subgroup of the AdS isometry group allows us to extend this to a set of compatible time dependent unitary flows along each of a discrete set of time-like geodesics, which lie at the centers of closed packed spheres on a hyperbolic slice of AdS.  The unitary embedding on a particular diamond is combined with those of a set of non-overlapping diamonds whose holographic screens cover the spatial sphere at a fixed radius, to make a full unitary map on the Hilbert space of the lattice model at that radius.  

The time evolution defined by these unitary embeddings of finite dimensional Hilbert spaces is necessarily discrete, and clearly the discrete interval should be taken as the Planck scale.  All explicit examples of CFTs with Einstein-Hilbert duals have two or more large compact dimensions whose size is of order $R_{AdS}$.  Since these dimensions are "hidden" inside single nodes of the tensor network, the appropriate Planck scale is the Planck scale in the AdS dimensions, after compactification.  The relation between this discrete time evolution and the continuous bulk time evolutions defined in\cite{LL} is obscured by the fact that their analysis takes place in the strict $N = \infty$ limit.  More speculatively, it may be related to the hypothetical ``double scaled" limit discussed briefly below.  

The question we want to address here is what connection, if any, there is between the finite area causal diamonds defined by the tensor network cutoff, and the ``infinite N finite causal diamond" of\cite{LL}.  In order to avoid confusion between two different notions of Type $III_1$ algebra, the authors of\cite{LL} suggest viewing their arguments as taking place in a cut off version of the CFT. However, in the mind of the present author this in fact generates some confusion.  The cutoff version of CFT that preserves the most structure of the $AdS_d/CFT$ correspondence is the tensor network/error correcting code framework.  As we have said, this is a sequence of lattice approximations to the CFT, laid out on spherical shells of a Riemannian hyperbolic space of dimension $d - 1$.  The Hamiltonian of the lattice models converges to a particular $K_0 + P_0$ generator of the conformal group in the infinite radius limit.  This is also the case for the time dependent Hamiltonians of the TNRG.  In the latter case, for simple models, it has been shown that the lowest eigenvalues of $K_0 + P_0$ are approximated well by those of the time dependent Hamiltonians of the TNRG on small lattices, implementing Maldacena's scale radius duality.  This is not true for the time independent Hamiltonians of the lattice models. For finite $N$ there are no sharp causal diamonds in the flow defined by the time independent Hamiltonian of any finite radius lattice model.  

However, as noted above, Fischler and the present author\cite{tbwfads1} pointed out that the tensor network renormalization group embedding maps of Evenbly and Vidal\cite{tnrg} did provide a notion of sharp causal diamonds in the tensor network formalism.  The time dependent Hamiltonian of the TNRG also converges to the $K_0 + P_0$ generator in the infinite radius limit.  Both of these approaches only define causal diamonds whose size is of order the AdS radius or larger, and are basically exploiting the redshift of energies between the interior and boundary of AdS.   A proper time of order $R_{AdS}$ at a radius of order $R_{AdS}$ is a proper time of order $r$ when $r \gg R_{AdS}$.  The observation of\cite{LL} is that when $\frac{R_{AdS}}{L_P} \rightarrow \infty$, (in the strong coupling regime where there are no other microscopic length scales), the boundary time evolution of single trace operators vanishes to leading order in $N$.  What this means is that the time scale for evolution is going to infinity like a power of $R_{AdS}/L_P$\footnote{The precise power depends on the particular holographic CFT under study.}. This is a point where I find myself a little confused because the arguments given in\cite{LL} for the Type $III_1$ nature of various boundary algebras always rely in the end on the standard $N = \infty$ map of single trace boundary operators into local quantum fields in AdS\cite{BDHM}\cite{HKLL}. 

If we take the infinite $N$ limit of a lattice model at a finite shell of a tensor network, each node of the network has an infinite dimensional operator algebra of "single trace" operators.  Our experience with the Type $0B$ model suggests that this will be a Type $I_{\infty}$ algebra, and since the lattice is finite, the same will be true of the single trace operator algebra of the full lattice model.  The scaling arguments of\cite{LL} remain valid and the algebra of operators in a (properly normalized) finite time interval will have a non-trivial commutant in the full operator algebra.  

In order to justify the identification of this commutant with the algebra of a bulk causal diamond, we have to make some further assumptions.  The first of these is that the Hamiltonian of the TNRG is a good approximation to the scale transformation $K_0 + P_0$.  The second is that what we mean by the algebra of "smeared single trace operators" in the lattice model is just the algebra of local gauge invariant operators at a lattice point.  The final assumption is that a precise notion of sharp causal diamond {\it does} exist in finite tensor networks and that it coincides with the diamonds defined in\cite{LL} in the $N = \infty$ limit, but persists even at finite $N$.  The key to this definition of diamonds is to search from the beginning for the analog of half sided modular flow, rather than try to define it in terms of a boundary time evolution operator.  A clue to this is that the ``diamond universe coordinate systems" defined inside causal diamonds by modular flows in conformal field theory\cite{CHM}\cite{JV} define a time dependent evolution operator even if the space-time has global time-like Killing vectors.   The time dependence of the evolution operator is necessary because if we choose any trajectory connecting the past and future tips of the diamond, and any nested sequence of proper time intervals, then causality requires that new degrees of freedom are encountered as the interval gets larger.  Thus, the Hamiltonian must change to include them.  The embedding maps of the TNRG are just such a structure.  

Given these assumptions we can claim that the picture of\cite{LL} coincides with the causal diamonds defined by the TNRG in the sense of $1/N$ perturbation theory.  A given interval of time in AdS radii corresponds to penetrating only a certain number of layers into the tensor network, in the following sense.  Although to a bulk observer a local node of the network near the origin has a size of order $R_{AdS}$, when it is translated into a boundary operator by the network it is spread over a finite fraction of the boundary, which contains a large number of nodes.  Therefore, its commutator with any local single trace boundary operator is down by powers of $1/N$.  In the $N = \infty$ limit, it is in the commutant.  

It is clear that these arguments only make sense in the limit that the tensor network is large, so that it forms a reasonable approximation to the geometry of the hyperbolic slice of AdS space.  Certainly they fail if the network consists of only a single node. This suggests that the arguments of\cite{LL} are valid in some kind of double scaled limit in which the UV cutoff length scale of the boundary field theory goes to zero like a power of $N$.   In $1/N$ perturbation theory they may be valid for fixed small boundary lattice spacing if the Lieb-Robinson tails of commutators outside the ``light-cone" come with inverse powers of $1/N$.

A final remark to avoid confusion: the tensor network provides a bulk UV cutoff of AdS space.  This has nothing to do with the boundary UV cutoff, which should be thought of as changing according to TNRG flow as one goes from shell to shell of the network.  When we say we're working with a fixed but large boundary UV cutoff, this refers to the number of nodes of the largest shell of the network.  The UV cutoff length scale in the bulk is {\it always} $R_{AdS}$, independent of the boundary cutoff.  In a CFT with an Einstein Hilbert dual this is always $\gg$ any microscopic length scale.  

\subsection{Comment on Type $II_1$ Algebras For Finite Area Causal Diamonds}

Following the work of\cite{LL} several papers appeared\cite{CLPW}\cite{jss}  suggesting that Type $II_1$ algebras were an appropriate description for finite area causal diamonds. If we accept the longstanding wisdom\cite{sw} that finite area diamonds can be described in a cutoff CFT, that one possible cutoff is a lattice, and the prescription of\cite{tbwfads1} that the holographic screens of finite area diamonds are the shells of tensor network sequences of lattice approximations, then the Type $II_1$ prescription is clearly wrong for any finite $N$.  The Hilbert spaces are all finite dimensional and algebras are all Type $I_M$ for some finite $M$.  This does not of course mean that there is {\it no} cutoff scheme for which the Type $II_1$ idea makes sense, but it does mean that any discrepancy between results in the Type $II_1$ definition and the finite dimensional TNRG definition of causal diamonds is non-universal and cannot be determined from CFT data alone.  When this observation is combined with the result\cite{tbsober} that most of the finite quantum information in a finite area diamond cannot, even in principle, be be accessed by any local measuring device inside the diamond, one concludes that the Type $II_1$ framework for describing finite entropy causal diamonds is unlikely to be very useful.  

\subsection{Cutoffs in the Type $0B$ String}

As we have noted, typical AdS/CFT models have at least two dimensionless parameters, the ratio of the AdS radius to the Planck length, which LL identify as a power of $N$, and the ratio of the string tension to the inverse square of the Planck length, which is $g_s^2$.  The Type $0B$ string has only one such parameter, $g_s^2$.  We would like to claim that the analog of $N$ in these $1 + 1$ dimensional models is the parameter $M$ that we have introduced to produce a system with meta-stable bound states whose properties resemble those of linear dilaton black holes.  Note that, at infinite $M$, bound states containing a finite fraction of the fermion flavors never decay away, despite the fact that the Hawking temperature is independent of $M$.  At any finite time there is always an infinite entropy remnant left over, because the time for black hole decay at finite $M$ scales like $M$.  

Recalling that the tensor network regularization of the CFT corresponded to an IR cutoff on AdS space, it's clear that the analogous way to impose an IR cutoff on the $1 + 1$ dimensional models is to back away from the double scaled limit and consider a finite number $N$, of fermionic particles, of $M$ different flavors, living in a double welled potential, with quartic interactions of the type we have proposed, concentrated at the origin.  The analog of the tensor network construction is a sequence of such models in which $N$ is increased by $1$ at each step and the distance between the two wells is increased at a rate corresponding to the double scaled formula for the fermi level.  The details of this will of course depend on the exact form of the double well potential.  The TNRG is a unitary embedding map of the Hilbert space of the $N$-th model into the $N +$ first.  If we turn off the interactions, it is clear how to structure this ``TNRG" so that it mirrors the space-time structure of the model.  Recall that the mapping between non-relativistic and relativistic fermions has two features.  First, we have the AKK map that directly converts the non-relativistic $\psi (\lambda)$ field into two species of Dirac fermions on a half line.  Then, we have a computation of the entropy of fermions confined to a region in $\lambda$\cite{dasetal}\cite{lindil1}, which, in the regime of strong string coupling, maps into the linear dilaton field as a function of the relativistic coordinate.  Thus, the key feature that the TNRG map has to preserve is the entropy formula as a function of $\lambda$.  We can do this by simply making sure that the low lying levels of the model with $N$ fermions are close to those of the model with $N + 1$ fermions.  It's possible that this requires us to make the potential $N$ dependent, but this is entirely in keeping with the spirit of renormalization.  One changes the parameters with scale, in order to keep the physics unchanged.  

Now we have a situation completely analogous to that described above for AdS/CFT models.  The system with an IR cutoff has a globally defined time independent Hamiltonian, with no sharp light cones.  However we can use the embedding maps of our ``TNRG" to define sharp causal diamonds.  These definitions are not unique, but they can be made to approximate the sharp causal structure of the double scaled models with arbitrary precision.  This last statement is special to two dimensional models.  In higher dimensions the exact models of quantum gravity defined by AdS/CFT, the large N limit of BFSS\cite{bfss} matrix models, and hypothetical resummations of string perturbation theory, do not have uniquely defined sharp causal structures away from the $N = \infty$ limit.  Tensor networks define sharp causal structures\cite{tbwfads1} in AdS/CFT for finite $N$, but they have the usual ambiguities of lattice regularizations.  The Hilbert Bundle formulation of Holographic Space-Time\cite{hilbertbundles} is an attempt to generalize tensor networks to length scales smaller than the AdS radius and to space-times with non-negative cosmological constant.  It has similar ambiguities.  It also illustrates that most of the quantum information in a finite area causal diamond is not accessible to measurement by a local detector in that diamond\cite{tbsober} so that no experiment can ever be a complete test of a model of the local physics of quantum gravity.  The fundamental reason for this is that most of the q-bits in a diamond are coupled together by a fast scrambling Hamiltonian, which transports information around the entire horizon in a way that no detector can follow.  

The other special feature of two dimensional models is that they always contain a preferred time-like geodesic.  In these models, entropy is encoded in a scalar field, and the gradient of that field is space-like in the asymptotic region\footnote{An exception to this is JT de Sitter gravity.  I believe the physics of this model has not yet been properly understood.}.  There are special time-like geodesics that go through the points of minimal and maximal entropy and it is usually enough to build a quantum evolution operator based on one of those geodesics.  

\section{Conclusions}

There are several conclusions to be drawn from our considerations.  The first is perhaps that $1 + 1$ dimensional models are special.  There are infinite entropy models of two dimensional quantum gravity that are mathematically equivalent to quantum field theories without gravity.  The UV infinities of the field theory are secretly IR infinities.  Thus, in this context at least the appearance of Type $III_1$ operator algebras has little to do with black hole horizons.  The more important lesson to be learned though is that sharp horizons can be imposed by time dependent Hamiltonians, and that the key to understanding them is the entropy formula relating the ``area" of a diamond boundary to the size of its quantum Hilbert space\footnote{I believe that for non-negative c.c. or diamonds smaller than the AdS radius for negative c.c., the correct connection is give by the Carlip-Solodukhin formula relating area to central charge of a near horizon conformal field theory, but that is not the subject of this paper.}. A sequence of embedding operators mapping smaller into larger Hilbert spaces is the quantum analog of a nested sequence of causal diamonds, which is the same as a time like geodesic.  A quantum theory of gravity in a given space time is a non isometric embedding of a whole collection of such nestings into a single Hilbert space.  This requires a set of compatibility conditions that ensure that quantum information accessible in two different nestings, which appears geometrically in the overlap between causal diamonds, is the same from the point of view of both sets of embedding operators.

In summary, the Type $III_1$ algebras that appear in $1 + 1$ dimensional gravitational models, although they appear to be related to bulk causal structure, are really more like the boundary subalgebras in AdS/CFT.  They encode the fact that the boundary Hilbert space is infinite dimensional.  When we impose an IR cut off, to mimic the origin of these models as the throats of finite entropy four dimensional charged black holes, the Hilbert space becomes effectively finite dimensional.  Bulk causal structure can be imposed on these finite dimensional systems by introducing the analog of half sided modular flows: time dependent embedding maps of lower dimensional Hilbert spaces into higher dimensional ones, guided by the entropy function of the classical gravitational solution.  These resemble the TNRG of AdS/CFT models, but $ 1 + 1$ dimensional static backgrounds always have a preferred minimal entropy geodesic, so a ``network" description is not really necessary.   The causal flows in finite entropy systems are necessarily unitary embeddings rather than full unitary maps on the full quantum Hilbert space, which is defined by the largest dimension space encountered as proper time gets large.  This of course implies that causal time evolution is discrete and time dependent, even when the background space-time has a global time-like Killing vector.  To complete it into a full unitary unitary transformation, we must construct a Hilbert bundle over the space of all time-like geodesics, with a (flat?) connection defined by the requirement that the entanglement spectrum of the density matrix on the maximal causal diamond in the overlap between two diamonds is independent of the fiber of the bundle in which it is computed\cite{hilbertbundles}.   

Although these general ideas apply to arbitrary background solutions of Einstein's equations, one can probably make the most rapid progress by following them through in the AdS/CFT context.  Here it would seem that the two most urgent projects are constructing higher dimensional tensor networks invariant under discrete subgroups of the isometry group of a global spatial slice of AdS, and an elaboration of the idea of a double scaling limit in which $N$ and the boundary UV cutoff are taken to infinity in some correlated manner.  

\vskip.3in
\begin{center}
{\bf Acknowledgments }
\end{center}
 The work of T.B. was supported by the Department of Energy under grant DE-SC0010008. Rutgers Project 833012.  Conversations with S. Leutheusser and H. Liu are gratefully acknowledged.




\end{document}